\documentclass[11pt]{article}
\usepackage[dvips]{graphicx}
\usepackage[utf8]{inputenc}
\usepackage[english]{babel}
\usepackage[T1]{fontenc}

\usepackage[title,titletoc]{appendix}
\usepackage{graphicx}
\usepackage{color}
\usepackage{float}
\usepackage{tabularx}
\usepackage{indentfirst}
\usepackage{hyperref}
\usepackage{pdfpages}
\usepackage{fancybox}
\usepackage{setspace}

\usepackage[T1]{fontenc}

\usepackage[table,dvipsnames,xcdraw]{xcolor}  
\usepackage{tabularx}
\usepackage{booktabs}
\usepackage{enumitem}

\usepackage{pdflscape}  
\usepackage{afterpage} 
\usepackage{float}  
\usepackage{placeins} 

\usepackage{graphicx}
\usepackage{pgfgantt}
\usepackage{xcolor}
\usepackage[normalem]{ulem} 
\usepackage{color,soul} 

\usepackage{multicol}
\usepackage{graphicx}

\usepackage{subfigure}


\newenvironment{capalayout}{
    \setlength{\topmargin}{-3.0cm}
    \setlength{\footskip}{0cm}
    \thispagestyle{empty}
} 

\newcommand{\ISSNno}{0103-9741}
\newcommand{\MCCSeqAno}{03/2023}
\newcommand{\TituloCapa}{Extracting Blockchain Concepts from Text}

\newcommand{\AutorANome}{Rodrigo Veiga}
\newcommand{\AutorAemail}{rodrigoveiga1206@gmail.com}
\newcommand{\AutorBNome}{Valeria de Paiva }
\newcommand{\AutorBemail}{valeria@topos.institute}
\newcommand{\AutorCNome}{Markus Endler}
\newcommand{\AutorCemail}{endler@inf.puc-rio.br}

\begin{document}

\begin{capalayout}
        \centering
        \includegraphics[keepaspectratio,width=14.7cm]{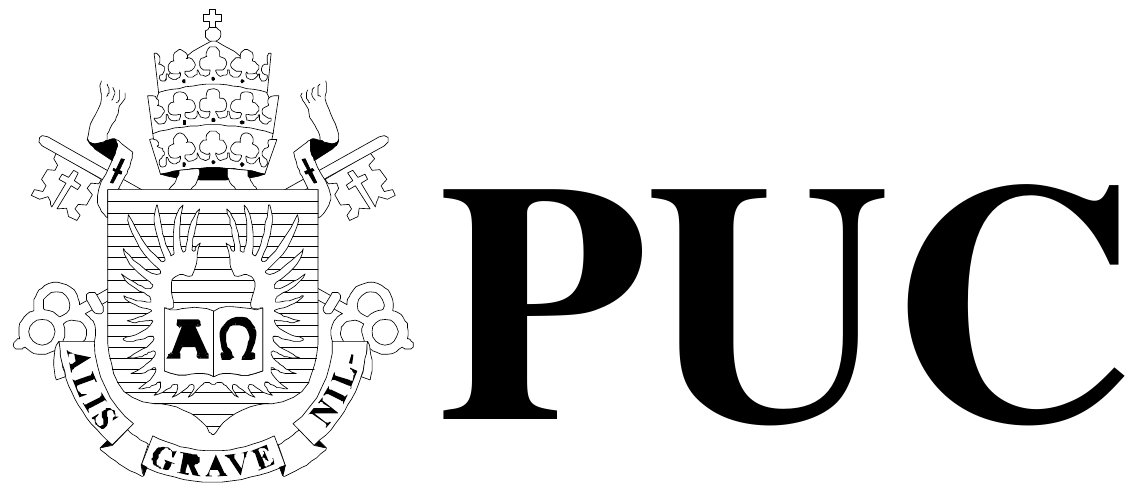}
        
        \medskip
    
        \setlength\fboxsep{1pt}
        \shadowbox{\fbox{\begin{minipage}[h]{14.5cm}
            \begin{center}
                \doublespacing
                \fontfamily{phv} \fontsize{14}{16} \selectfont
                \medskip
                ISSN \ISSNno
                
                \bigskip
                Monografias em Ciência da Computação 
                
                n\textordmasculine \, \MCCSeqAno
                
                \bigskip
                \medskip
                \fontsize{18}{20}\selectfont
                \textbf{\TituloCapa}
                
                \bigskip
                \fontsize{14}{15}\selectfont
                \textbf{\AutorANome} \\
                \textbf{\AutorBNome} \\
                \textbf{\AutorCNome}
                
                \bigskip
                
                \medskip
                Departamento de Informática
                \bigskip
            \end{center}
        \end{minipage}}}
        
        \bigskip
        \bigskip
        
        \begin{minipage}[h]{14.5cm}
            \doublespacing
            \fontfamily{phv}
            \begin{center} 
                \fontsize{12}{14}\selectfont
                \textbf{PONTIFÍCIA UNIVERSIDADE CATÓLICA DO RIO DE JANEIRO} \\
                \textbf{RUA MARQUÊS DE SÃO VICENTE, 225 - CEP 22451-900} \\
                \textbf{RIO DE JANEIRO - BRASIL}
            \end{center}
        \end{minipage}
        \newpage
\end{capalayout}
\thispagestyle{empty}

\begin{flushleft}
    \begin{tabular}{p{11.1cm}r}
        Monografias em Ciência da Computação, No. \MCCSeqAno & ISSN: \ISSNno \\
        Editor: Prof. Editor Principal & Mar, 2023
    \end{tabular}
\end{flushleft}
\LARGE
\bigskip

\begin{center}
    {\bf \TituloCapa}
\end{center}
\normalsize
\bigskip

\begin{center}
    {\bf \AutorANome, \AutorBNome and \AutorCNome}
\end{center}

\begin{center}
   \AutorAemail, \AutorBemail, \AutorCemail
\end{center}


\noindent {\bf Abstract.} Blockchains provide a mechanism through which mutually distrustful remote parties  can reach consensus on the state of a ledger of information.  With the great acceleration with which this space is developed, the demand for those seeking to learn about blockchain also grows. Being a technical subject, it can be quite intimidating to start learning. For this reason, the main objective of this project was to apply machine learning models to extract information from whitepapers and academic articles focused on the blockchain area to organize this information and aid users to navigate the space.
\medskip

\medskip


\bigskip
\noindent {\bf Resumo.} 
Blockchains produzem um mecanismo através do qual partes distintas que nao confiam mutuamente umas nas outras podem chegar a um consenso sobre o estado de um registro de informações. Com a grande aceleração com que o espaço blockchain se desenvolve, cresce também a procura de informacoes por parte de quem busca aprender sobre blockchains. Sendo um assunto bastante técnico, esse pode ser bastante intimidante para quem quer começar a aprender. Portanto, o objetivo principal deste projeto foi aplicar modelos de aprendizado de máquina para extrair informações de `whitepapers' e artigos acadêmicos focados na área de blockchain para organizar essas informações e auxiliar os usuários a navegarem esse espaço.
\medskip
\medskip

\noindent {\bf Palavras-chave:} blockchain; aprendizado de máquina; extração de informação \\
\newpage
\pagenumbering{roman} \setcounter{page}{2}
\vspace*{\fill}
\begin{flushleft}
    \textbf{In charge of publications:} \\
    PUC-Rio Departamento de Informática - Publicações \\
    Rua Marquês de São Vicente, 225 - Gávea \\
    22453-900 Rio de Janeiro RJ Brasil \\
    Tel. +55 21 3527-1516 Fax: +55 21 3527-1530 \\
    E-mail: publicar@inf.puc-rio.br \\
    Web site: http://bib-di.inf.puc-rio.br/techreports/ \\
    \end{flushleft}

\newpage


\pagenumbering{arabic} \setcounter{page}{1}
\title{Extracting Blockchain Concepts from Text}
\author{Rodrigo Veiga \and Markus Endler \and Valeria de Paiva }
\date{March 31, 2023}

\maketitle
\section{Introduction}
Blockchains provide a mechanism through which mutually distrustful remote parties (the nodes) can reach consensus on the state of a ledger of information. Blockchain  concepts were first introduced   with the aim of implementing a system where information about the creation of a document was immutable~\cite{Haber1991, sherman2019}. 
It took almost two decades, with the invention of Bitcoin, for this concept to be part of a real application~\cite{conway}.

A blockchain is an approach of Distributed Ledger Technology (DLT), which is a transaction log based on blockchains that contain information about various transactions. 
These blocks maintain their integrity through a collaborative network of computers working to read and validate them. The computational work carried out relies on cryptographic mathematical methods to keep the network safe and reliable for each user's data. 
Any computer can connect to the network as a new participating node and perform tasks to validate blocks of transactions. Upon being validated by one of the nodes, a block is then added to the blockchain and the remaining nodes of the network start to see this new extended chain as the source of truth. In this way, the nodes participating in the network become a central piece in the functioning of a blockchain.

This brief description of the blockchain is taken from the article that introduced its first successful application to scale, the original Bitcoin whitepaper~\cite{nakamoto2008}. This file was released on January 4, 2009 via mailing lists to crypto enthusiasts by an unknown entity using the pseudonym Satoshi Nakamoto. With its release Bitcoin became the first successful attempt to create a system of electronic currency fully digital and decentralized. 

The state of the world news context in 2008 helps us  understand part of the motivation behind this invention, as the world was trying to recover from one of the biggest financial crises in history at that time. A major factor that contributed to the financial collapse was the irresponsibility of some banks and large financial institutions, which for decades were considered pillars of the modern economy. These acted as a source of trust in the financial market but then found themselves in the opposite position, being harmful to the financial lives of their clients and a reason for great distrust. It really seemed like the right time to bring money to the internet, and so it was done. After 15 years, the Bitcoin network (\url{https://bitcoin.org/en/}) continues to fulfill its main objective: to use a consensus mechanism to guarantee the veracity and integrity~\footnote{The biggest threat to blockchains is the existence of "51\% attacks", which may occur when a malicious actor controls more than 50\% of all computers connected as nodes on the network (\url{https://dci.mit.edu/51-attacks}). 
} of transactions carried out on its network.

While many people view this technology with a degree of distrust, as is the case with any innovative technology, blockchain and Bitcoin have clearly gained a lot of popularity and adoption in recent years, especially among generations under 40-years-old, and now internet natives as well. This shouldn't surprise us, given the fact that this audience is more dependent on and more used to computers and the internet. For many of these people, a form of money native to the internet seems completely natural, and some are even surprised that none has been fully adopted yet.

\section{Digital currencies}
Financial services are increasingly accessible to anyone with a mobile device. Nearly 50\% of the global population is already using their cell phones to at least perform balance and statement inquiries, according to the Nielsen Group in 2016. 
With more and more devices with internet access and connected to each other, we see great promises of innovation in the financial sector, and one of them is the use of blockchain. For devices with internet access and connected to each other, new innovations would become possible if they had the ability to carry out financial transactions with each other. These transactions are not just limited to cell phones, but to any electronic device that can act as a participant in the economy. A classic example of a blockchain application on electronic devices is a vending machine programmed with a "smart contract" to take payments and release products as soon as they are paid for. These  smart contracts are one of the main innovations of  blockchain. They are defined as a kind of bank account not controlled by a user, being a program implemented in a blockchain network that works according to the rules established in the code, or in the contract. These contracts have their own balance and are able to receive and carry out transactions, enabling the user to interact with them.

In view of this scenario of increasingly connected devices, interacting with both humans and other machines, the invention of  blockchain could be the catalytic event for the financial sector to fully adapt itself to the digital environment, thus becoming another key technology in the gears of “Industry 4.0”. Industry 4.0 is the name given to the new industrial revolution that involves intelligent and interconnected means of production and distribution, highly dependent on information obtained in real time~\cite{Javaid2021}.

\subsection{Exponential growth?}
Bitcoin's source code was immediately made available to anyone who wanted to analyze or improve it. Being an open source software, in the following years the Bitcoin and blockchain community tried to improve the technology similarly to how the Linux community transformed the operating system into one of the best known and most successful in the world see \cite{infoworld2009}. Information and development are shared by thousands of creative minds, and this has led to the creation of diverse projects and technologies that can work together in order to reach innovative end goals, creating new solutions to previous problems.

Regarding Bitcoin, there are still important technical issues to be solved, with the biggest one probably being the expensive fees and slow transfer speed when the network is congested, which it usually is. For this reason, it is easy to see that Bitcoin does not serve as an electronic money unit to the point of replacing  the dollar (or the Brazilian real) for daily use. This type of technical problem is complex and difficult to solve, but probably achievable. The probability of this problem having a solution becomes even greater due to the fact that there are thousands of people looking for solutions every day - after all, everything is open for anyone to participate. 

The fact that anyone can create a copy of Bitcoin software has allowed for the creation of major blockchain innovations, with the most notable example, perhaps, being Ethereum, a cryptocurrency proposed in 2013 and released in 2015 by then-Russian-Canadian programmer, 21-years-old Vitalik Buterin \cite{finley2014}. Unlike Bitcoin, Ethereum is a Turing-complete blockchain, giving this network more flexibility and allowing the creation of smart contracts, which can be programmed similarly to modern programming languages. 

The introduction of smart contracts was a real milestone in the blockchain field - with it, transferring money is no longer the only possible application. A concept that has gained traction recently is decentralized finance (DeFi), which makes use of smart contracts to replicate traditional financial services (such as loans, insurance and derivatives) in an open, transparent and interoperable way \cite{schar}.

Currently, there are more than 9,000 different cryptocurrencies according to \url{https://coinmarketcap.com/}, and while the vast majority of them are negligible, many have incredibly active teams and communities focused on developing innovative projects. In fact, new technologies are introduced so quickly that it can be quite cumbersome, if not impossible, to keep up with everything that's going on. To truly understand the value of many of these projects, it is necessary to possess not only the technical skills but also the willpower to do some in-depth research. Even meeting these prerequisites, it is common to need to go back to a previous topic to understand a new one. If there's any downside to collaborative and open-source development on this scale, it must be this constant need to keep learning so you don't get left behind. In most cases, it is difficult to even know where to start. While many sites do a good job listing the top projects, they don't give us good insights into what these projects do and how they relate to each other. This idea of linking projects may not make much sense in the traditional business world, but it is very important in the blockchain area and can be crucial to understand this field, as these projects are always in constant collaboration.

\subsection{Motivation and objectives}
With the great acceleration with which this space is developed, the demand for those seeking to learn about blockchain also grows. Being a somewhat technical subject, it can be quite intimidating to start learning, and many people don't even know where to start. For this reason, the main objective of this project was to apply machine learning models to extract information from whitepapers and academic articles focused on the blockchain area to organize this information and aid users navigate the space.

The prediction models we use were trained with the SciERC dataset, created by Yi Luan et al~\cite{luan2018}, which was constructed from information taken from computer science articles. We seek to obtain the entities and relationships taken from blockchain texts in order to generate a knowledge graph that can be used for research and learning. The realization of this work can be seen in four parts:
\medskip

\noindent {\textbf{Collection of quality data related to relevant projects and technologies in the blockchain ecosystem, separated into two distinct corpora}}
\begin{enumerate}
\item
No blockchain-related dataset was found on the internet that was suitable for the purposes of this work. We only found websites with links to whitepapers and academic articles. It is necessary to extract data from these documents and group them as follows:

\item 
Corpus formed by phrases from whitepapers published by relevant projects in the blockchain area. Whitepapers are similar to scientific articles in that they are very technical and describe a technology to address a problem. Although technical in contents, they are not evaluated and judged by other researchers, as is the case with articles published in scientific conferences and scientific journals.

\item 
Corpus formed by sentences from academic articles related to blockchain.
The creation of the corpus was one of two products generated by this work, and will be exposed to the public that wants to use it in future research.
\end{enumerate}

\noindent{\textbf{Execution of DyGIE++ framework to train the prediction model and apply it to the collected data}}

\begin{enumerate}
\item Using the SciERC dataset to train the
prediction. SciERC is a collection of 500 scientific abstracts, with language similar to that of the collected data.

\item Analysis of results

\item Comparison of entities and relationships obtained by the model with the result of entities and relationships obtained manually
\end{enumerate}

\noindent{\textbf{Development of dygiepp-reader to format results and feed the web interface}}

\begin{enumerate}
\item Creation of the dygiepp-reader Javascript program [14] to
make the results available to a web interface
\end{enumerate}

\noindent{\textbf{Development of a web interface for viewing results}}

\begin{enumerate}
\item Create a web interface for viewing and interacting with results

\item This corresponds to the second product generated by this project, it will be made available on the internet in the public domain for those who wish to use the tool for research or learning purposes.
\end{enumerate}

As blockchain projects and technologies are in constant elaboration and improvement, it is very important to understand how these projects relate to each other to understand exactly what is happening in the area. As there are currently no tools that allow this type of visualization, we imagine that there are important technological opportunities that we have yet to discover when we have an enlarged view of the blockchain ecosystem. 

We will use the SciERC dataset to train a prediction model, and thus we will be able to generate a semantic graph with entities (concepts, problems, methods, etc.), and the relationships that exist between them (used-for, functionality-of, etc). This prediction model will be trained, evaluated and executed using DyGIE++, a Python framework developed by authors of the paper that introduces SciERC~\cite{luan-etal-2019-general}.

Bearing in mind that no online resources were found to serve as a dataset for generating a knowledge graph and the growing demand for information related to blockchain technologies and projects, this project seeks to tackle these problems with the development of two final products.

The first is a corpus of text files on blockchain-related technologies, concepts and projects that is large enough to build a knowledge graph. 
The second is a web interface provided on the domain that allows visualization and interaction with this graph, linking concepts, or entities, to each other through relationships. We hope not only to make life easier for those wanting to learn about blockchain and cryptocurrencies, but also to help further research with the collected data, which will be made public and posted on blockchain forums for feedback and free use.

\section{Tools for KG construction}
We describe the tools we will use in  this project to extract blockchain concepts from text and then organize these concepts into knowledge graphs.

\subsection{AllenNLP}
With the advancement of scientific research in any academic area, the analysis of information becomes more complex due to the high volume of published articles, making information sources increasingly dispersed~\cite{dessi2020}.

Fortunately, the rapid development of natural language processing (NLP) tools allowed the construction of knowledge graphs of a specific domain, containing its entities and linking them through relations \cite{Wang_2020}. In this way, the information is organized and interconnected, in a way that facilitates research and scientific development.

One such tool is AllenNLP, a library that applies deep learning models to NLP tasks, providing commands via the command line that are accessible to users who are not very familiar with NLP applications, which would often require specialized code development. This advanced knowledge is beyond the scope of graduation, and as this project involves the creation of blockchain-related knowledge graphs, it would not be possible to realize it if it weren't for this great tool that lowers the barriers to entry for NLP research, both for newcomers in this area as well as for more reputable researchers who live in environments of greater technological need \cite{gardner-etal-2018-allennlp}.

AllenNLP contributes to accelerating the amount of research and NLP experiments, and also has the advantage of being an open-source project, having its repository maintained by reputable engineers and researchers. Its repository currently has more than 10,000 stars and 2,000 forks, with its community growing at a fast pace and becoming one of the main NLP tools available.

\subsection{Entities and relationships}

It is possible to make sense of academic texts and articles using prediction models in AI, which detect entities and relationships in a set of sentences. These models are trained with annotated datasets with entities and relationships, ideally made by professionals and reputable people in the field about which a text is concerned, and the quality of the annotations is of vital importance for the success of data prediction. In this project we use the annotations of the SciERC dataset, a collection of 500 scientific abstracts \cite{luan-etal-2019-general}. These annotations contain the entities and relationships of the collected texts, and we will use the DyGIE++ framework to train a model that makes use of the SciERC dataset to predict entities and relationships in our collection of blockchain texts. With these results, we can create a blockchain knowledge graph and use it as a tool to research and analyze this knowledge network.

This work was inspired by the article by Yi Luan et al \cite{luan2018}. Their work has two main products: a framework called SciIE (whose acronym stands for extracting information from scientific files) for identifying entities, relationships and co-references in scientific articles; a dataset called SciERC composed of scientific articles with annotations on entities, relationships and co-references. The SciIE prediction model uses SciERC data that has been annotated by experts, containing their true entities and relationships. These annotations are important for training the model to predict entities and relationships with some accuracy. The final result of SciIE is a knowledge graph containing relationships between the entities detected in the data corpus. Graphs are first constructed at the document level, that is, having their scope limited to the sentences of only one document in the corpus of scientific articles. With these sub-graphs constructed, the entities and relations are joined into a single graph. 

In this graph, nodes correspond to entities that can be any of the following types:
\begin{itemize}
\item Tasks (applications, issues to be resolved)
\item Methods (methodology, models, tools used, etc.)
\item Evaluation Metrics (Metrics, measures that can be used to evaluate a method or system)
\item  Material (Data, resources)

\item Other Scientific Terms
(Scientific terms that do not fit into any other category)

 \item Generic  terms
 General terms or pronouns that may refer to an entity but are not themselves informative, often used as connective words.
 
 \end{itemize}

Edges represent the relationships between nodes or entities, and edges are one of the following:
\begin{itemize}
\item  Used-for 
\item  Feature-of 
\item  Hyponym-of 
\item  Part-of 
\item  Compare, entities being compared 
\item   Conjunction, entities being used together to reinforce the same point, see annotation guidelines in \url{http://nlp.cs.washington.edu/sciIE/annotation_guideline.pdf}.
\end{itemize}

Knowledge graphs are semantic networks, representing ideas and concepts and illustrating the relationship between them. In the case of SciIE, it was used to generate knowledge graphs related to the area of artificial intelligence, but it can be applied to any area of knowledge. With the knowledge graph, abstract ideas can be  visualized linking the concepts that work together to formulate the idea. In a new and rapidly developing area like blockchain, the construction of a semantic graph linking your main concepts and technologies can be useful for educational and research purposes.

\begin{figure}[h!]
\begin{center}
\includegraphics[width=10cm]{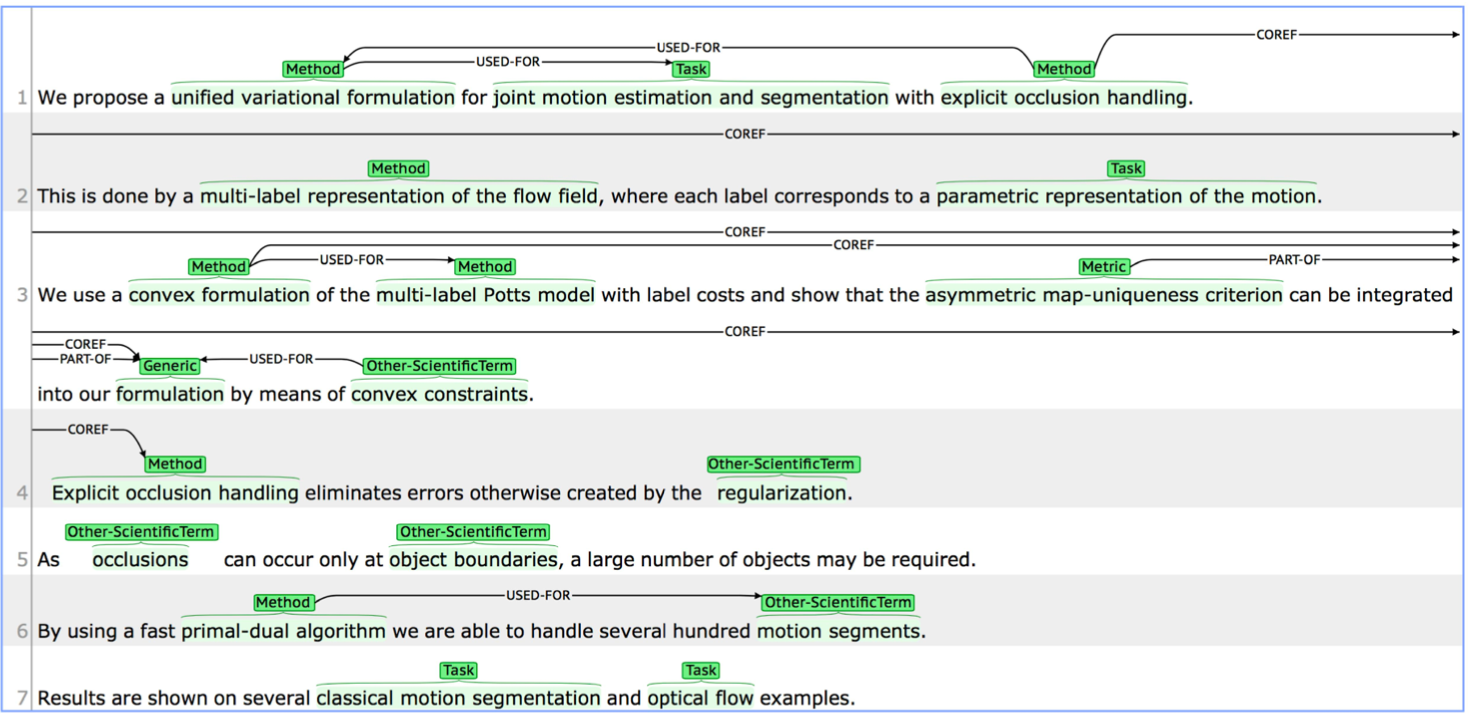}

\caption{Example of annotations used by the SciERC dataset}
\end{center}
\end{figure}
 
\subsection{SciERC}

SciERC is a collection of over 500 scientific abstracts annotated with their entities, relationships and co-references. These data were annotated by specialists in the area covered by the abstract in question. SciERC is an extension of the SemEval 2017 and SemEval 2018 datasets, and includes more entity types and relationships and phrase links using co-reference links \cite{luan2018}. SciERC's key point of interest for this project is that its articles are focused on the computing area, more specifically on artificial intelligence. As there is no annotated dataset with articles on blockchain, it was necessary to look for annotated data that resembles the type of language found in whitepapers and academic articles on blockchain, and SciERC was the nearest one that we could find.

\subsection{ DyGIE++}
In this work we use DyGIE++, a Python framework developed by Wadden et al \cite{wadden-etal-2019} that makes use of AllenNLP and its command-line interface to train, evaluate and apply prediction models for extracting entities and relationships in a data set. This program allows the model to be trained with 4 different datasets, including SciERC, our preferred dataset. DyGIE++ provides scripts to train and evaluate a model, and 
also perform predictions on a dataset. Although our project was inspired by the article by Yi Luan et al \cite{luan2018}, the DyGIE++ framework proved to be a quick gateway for the purposes of this work, in line with the accessibility that AllenNLP promises.

DyGIE++ has the scierc and scierc-lightweight models -- both use the SciERC dataset, but only the former makes use of coreferences. When the authors refer to scierc and scierc-lightweight (in lowercase), these are the prediction models used. The scierc and scierc-lightweight models were evaluated, with the first having 64\% accuracy for entity extraction and the second with 66\% accuracy. For the extraction of relations, scierc presented 44\% accuracy against 55\% for scierc-lightweight. Although the scierc-lightweight model initially gave a better result, both models will be discussed later. The scierc model differs from the scierc-lightweight because it makes use of coreferences, that is, it can associate words in different sentences, while the scierc-lightweight is limited to the scope of the sentence.

To perform predictions on a new dataset, it was necessary to format it the way DyGIE++ expects text. Fortunately, DyGIE++ provides a format-new-dataset script to format the data. This proved not to be enough, and the results only became adequate when we removed the line breaks from the data via a script created for this job. With the formatted data, we can then perform the prediction with the trained model, and thus generate the result files that represent the knowledge graphs, separated at the document level.

\subsection{Node.js and React}

In order to develop a web interface to visualize the final knowledge graph, it was necessary to develop a program that could manipulate the data from DyGIE++ result files and produce them in a more suitable way for a web page. DyGIE++ stores the results in a jsonl file, containing a list of objects in JSON format. We chose Node.js to accomplish this task due to the fact that Javascript is a language specifically designed for the web, and Node.js has a wide variety of libraries that allow the construction of APIs. The Express library was chosen for fast
 development of an API that provides a route to query the data obtained and display the knowledge network. The result was a program  called dygiepp-reader, and this will be covered in more detail in later sections.
 
To develop the web interface, React was used. React can be seen as one of the best solutions for implementing a website. To make calls to the dygiepp-reader API, I used the Axios library, which is also familiar.

\section{ Work development}

A corpus with 105 text files pulled from blockchain project whitepapers was chosen.
Then 57 text files were taken from academic articles related to blockchain.
We complemented it with 32 text files taken from Wikipedia pages related to
blockchain, for a total of 194 files. 

We note that we expected that there would be more academic articles detailing blockchain projects, the main focus of this project. 
But we observed that most articles sought to analyze specific technologies and concepts related to aspects of blockchain, not full projects.

\subsection{Using DyGIE++ to get entities and relationships in the dataset}
With the data collected, some steps are needed to generate the result file. First, it is necessary to train the prediction model on the desired dataset. In this case we want to train the model with SciERC, and we can do it with the following steps:
First, download the dataset.
\begin{itemize}
\item DyGIE++ provides a script to download all datasets it makes available, including SciERC
\item Then, train the model,
 using a script provided by DyGIE++ that accepts the name of the dataset used as an argument, in this case scierc.
 \end{itemize}

 Then run allennlp's `predict' command on the formatted data, specifying a file to store the results


\subsection{ Construction of the dygiepp-reader}

Using Node.js, the dygiepp-reader program was developed. This is responsible for structuring the DyGIE++ results in a more intelligible way. Using this program, we are able to perform the following tasks:
\begin{enumerate}
\item Get the terms that were most frequently detected as entities.
This listing was used to get a quick idea of the terms that were most detected as entities, serving as a reflection of the data that was collected. It was also used to highlight the most relevant entities in the web interface created to visualize the results.
\item Create an entity dictionary, where we can easily store information about each entity, such as:
\begin{enumerate}
\item[a.] How many times  has it been detected as an entity type?
\item[b.] In which sentences it was detected as an entity?
\item[c.] Which relationships is the entity a part of?
\end{enumerate}
\item Serving data to a web client via an API
Node.js was used with the Express library for the rapid development of an API so that web clients can consume the data obtained by dygiepp-reader. When making the call, the API responds with the entity dictionary and the list of the most frequent entities in the graph.
\end{enumerate}

The entity dictionary makes it easy to view the results during the
development, and most importantly, facilitates data access to create a visual knowledge graph. This kind of structuring is much nicer to the human eye as it allows us to see exactly which terms relate to each other, unlike how DyGIE++ generates the results in the jsonl file, using integers to represent the starting and ending position of the terms in a sentence.
 
This program has a class called \texttt{Visualizer}, which is initialized by receiving a path to the jsonl result file. In the construction of the class, a variable containing the JSONs in a list is initialized. 

For each line of the .jsonl file created, that is, for each JSON that
represents a document, we have a function which creates an entity dictionary, which will be important in the construction of the web interface to visualize the resulting graph. This dictionary allows direct access to the data of an entity, containing:
\begin{enumerate}
\item  The relationships the entity is part of, either on the left or right side of the relationship;
\item The sentences where the entity was detected as such;
\item A count for the number of times the term has been detected, as each entity type.
\end{enumerate}

This function receives as parameter a true or false value that specifies whether the use of {\textit{aliases}} should be used. If no value is passed, the function uses the aliases. An explanation of aliases and why they are important will be seen  later.

The function {\texttt{getFrequencyList}}
returns a list of all terms that were considered entities, ordered by the number of times each term was detected as such. This list takes into account all documents in the dataset, and each element in the list is a list in the format {\texttt{[entity, \# of times detected as entity]}}. This list serves two main purposes. The first is to
assess whether the results generated are aligned with the contents of the documents used in generating the resulting graph.
The second purpose is to  highlight the more frequent entities in the result graph visualization. 
The function {\texttt{ getGlossaryEntities}}
returns a dictionary containing all the glossary terms that were detected as entities and how many times were they detected.

\begin{figure}[!h]
\begin{center}
\includegraphics[width=7cm]{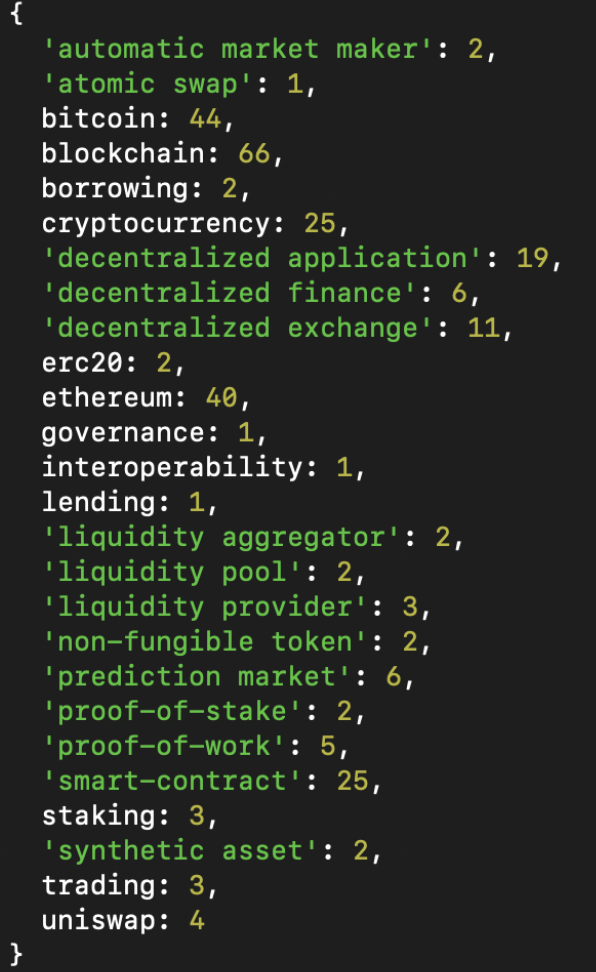}
\caption{Example of result of the function \texttt{getGlossaryEntries}}
\end{center}
\end{figure}


The program contains a directory called \texttt{lib}, where we have \texttt{aliases.js}, \texttt{glossary.js} and \texttt{visualizer.js} files. These first two are important for the evaluation and improvement of the results obtained, which we will see in more detail in the section on evaluating the  results. The \texttt{visualizer.js} contains the Visualizer class described above. It has a folder containing the scripts that run the analysis of the results (\texttt{analyze.js}) and transform a line from the jsonl file into a JSON given a document identifier (\texttt{get-json.js}). The jsonl result files are located in the data directory, inside the folder with the name of the prediction model used in the generation of the result file.

\subsection{Evaluation of the results obtained}
Having the results obtained from the prediction model, it is necessary to analyze them in order to know if they are good enough to serve as a basis for a knowledge graph that seeks to be a map of the blockchain ecosystem. For this, we want the prediction model to meet the following expectations:
 We would like to be able to detect entities and relationships with a hit percentage 
approaching 64\% and 55\%, which was the one described for the original experiment with SciERC.

In order to eyeball  the quality of the results obtained, we chose five
phrases from the whitepapers dataset and five phrases from the academic papers dataset and
manually checked and annotated their entities and relationships. With this  annotated data, a comparison was made with the results obtained by the DyGIE++ prediction model. The results obtained are shown in the following table.

\begin{table}[h!]
\centering\begin{center}
\begin{tabular}{ |c|c|c|c|c|c|c| } 
 \hline
 document & \# entities & \# right ent &\% correct &\# relns & \# right rels & \% corr relns \\
 \hline
acad$_2$ & 3 & 3 &100\% &1&1&100\% \\ 
acad$_3$ & 3 & 2& 66.6\% &2 &2&100\% \\ 
acad$_7$ & 3 & 2 &66.6\% &2 &2&100\%\\ 
acad$_8$ & 3 & 3 &100\% &1&0&0\%\\ 
acad$_{14}$ & 5 & 4 &80.00\% &5 &3&60\%\\
wp-net & 5 & 3 &60\% &1&0 &0\% \\ 
wp-chain & 2 & 2 &100\% &0&0&100\% \\ 
wp-polka & 2 & 2 &100\% &0&0&100\% \\ 
wp-uniswap & 9 & 7&77.8\%&0&0&100\% \\ 
wp-loop & 7 & 3&42.9\%&4&0&0\% \\ 
 \hline
\end{tabular}

\caption{Results of manual evaluation of entities and relations}
\end{center}
\end{table}

 
Taking the average of the correct results we obtain 73.8\% of correct entities and 50\% of correct relationships. We then see that for the entities we obtained a percentage of correctness greater than the 64\% given in the evaluation of the model. The percentage of correct answers for the relationships was a little lower than expected, but not enough for the work to be compromised.

An analysis was also carried out using Wikifier \url{http://wikifier.org/} wiki, a site that allows the user to send a text file and their verbs, nouns, adjectives and adverbs highlighted in different colors are returned. For this analysis, results of the prediction model in 10 sentences from the corpus of whitepapers was compared with Wikifer results.

The purpose of this analysis is focused more on entities - we want the nouns detected by Wikifier to also be detected as entities by DyGIE++. Remember that not all Wikifier nouns should be detected as entities, as entities are limited to the types described before. 
With this evaluation, we reached a hit percentage of 68.65\% for the entities, a value close to the 64\% that was obtained in the evaluation of the prediction model. With these two analyses performed, we would like to conclude that the DyGIE++ prediction model using SciERC data does not lose its accuracy when applied to the blockchain corpus dataset.

 We want it to be able to detect the entities relevant to the purpose of the project
Although the prediction model has detected entities and relationships with an acceptable degree of precision, we also need to know if the detected entities are the ones that interest us, that is, the ones we want to see in our knowledge graph. For this, it was necessary to create a glossary of terms of interest. In the dygiepp-reader program, a glossary.js file was created containing terms relevant to the interests of this project. Naturally, they are blockchain-related terms, with emphasis on DeFi, decentralized finance. There is a humanly compiled website (DeFi Pulse, \url{https://www.defipulse.com/}) that helped with the manual glossary construction.
The \texttt{getFrequencyList} function of the Visualizer class of dygiepp-reader was also used, which returns an ordered list of the most detected entities. This list served not only to find important terms missing from the glossary, but also to find entities related to a glossary term that could be considered synonyms or variations of the term. For the purposes of this paper, we refer to these entities as
\textit{aliases}. For this reason, an \texttt{aliases.js} file was also created, which contains a dictionary where each key is a glossary term and its value is a list of terms that should be considered an alias of this glossary word.

Aliases are important because they allow us to store the relationships and phrases of multiple entities with the same meaning in one place. A clear example is entities in the plural. As one example, we've added "smart contract" (no hyphen) and "smart contracts" to the list of aliases for "smart-contract", a glossary term. Whenever any detected entity is an alias, we add it to the entry in the entity dictionary that matches the glossary word it refers to.
With the glossary built, we can then know how many of its terms were detected as entities. We did an analysis of the percentage of terms detected with and without the use of aliases. For this, the \texttt{analyze.js} script was created, which calculates the percentage of glossary terms that were detected as entities for each corpus, and also counts the number of relationships detected for glossary terms. This script accepts as a parameter a boolean value indicating whether aliases should be used when initializing the entity dictionary. The analysis was performed with and without the use of aliases, for the results of the scierc and scierc-lightweight model. We report only the results for the scierc model in the following table:

\begin{table}[h!]
\centering
 \begin{tabular}{||c c c||} 
 \hline
 Corpus & Glossary Terms as entities & \# relations for glossary terms  \\ [0.5ex] 
 \hline\hline
 whitepapers &25 out of 47 terms detected (53\%) & 361 rels \\ 
 academic &17 out of 47 terms detected (36\%) &
155 rels \\
 wiki & 16 out of 47 terms detected (34\%) &
151 rels \\ [1ex] 
 \hline
 \end{tabular}
 \caption{Results for the model sciERC}
\end{table}

\subsection{Web interface for viewing results}

In the course of this work, dygiepp-reader evolved from a program intended only for use via the command line to a simple API that has a route for web clients to consume its data, such as the entity dictionary and the most frequent entities. React, a javascript library that facilitates the construction of web pages, was used to develop a website that makes calls to the dygiepp-reader API and displays its results. The site is still under development.
So far, the site has three sections: the section on the left displays buttons for the 100 most frequent entities; the middle section shows the relationships for this entity type (currently just the hyponym-of, part-of, functionality-of, and used-for relationships); the section on the right shows the phrase where a clicked relationship was detected. The three sections interact with one another -- clicking on an entity in the left section displays its relationships, and these can also be clicked to update the right section.

\begin{figure}[h!]
\begin{center}
\includegraphics[width=7cm]{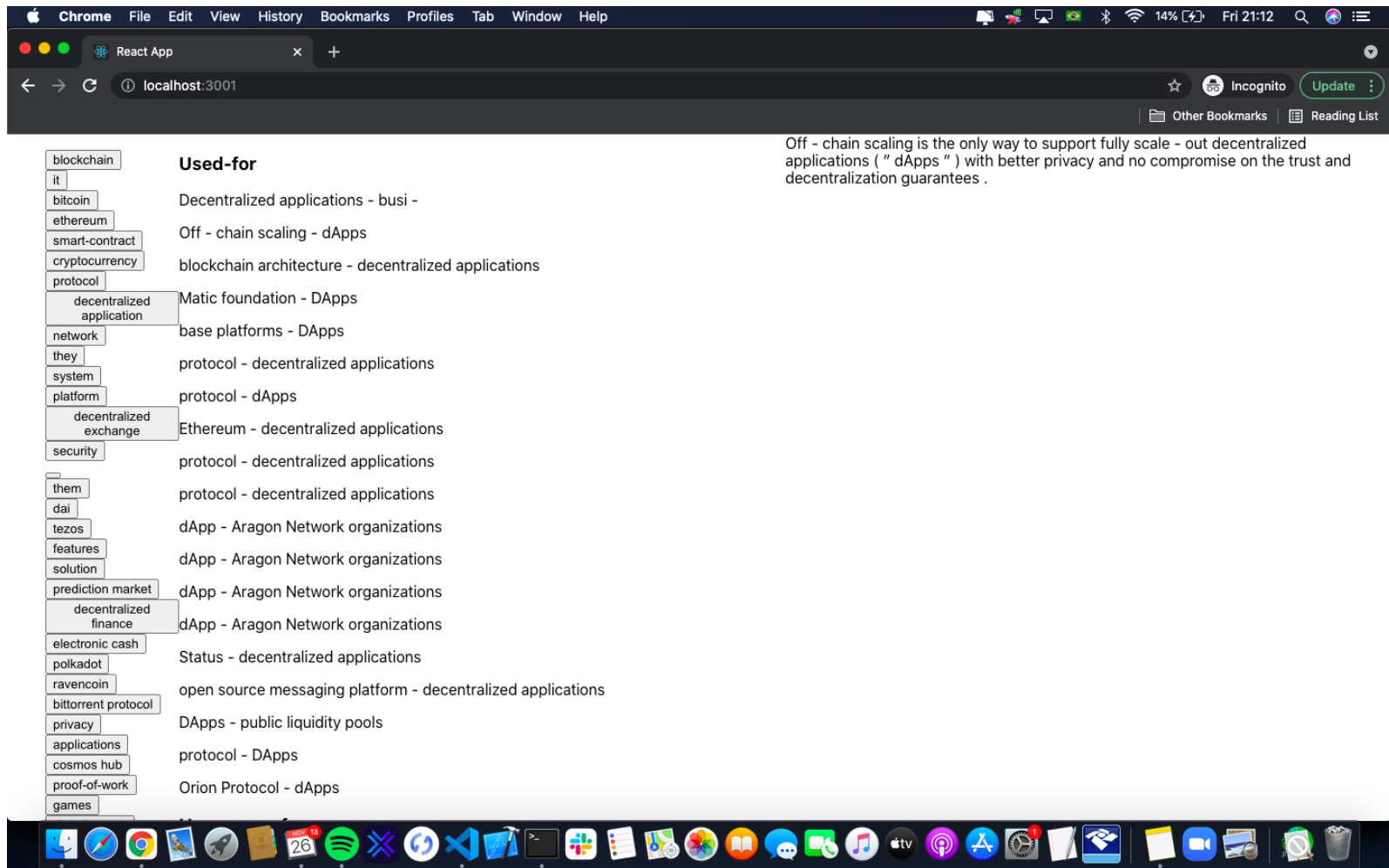}\\
\caption{Website visualizing the relationship `off-chain scaling--dApps' (Used-for)}
\end{center}
\end{figure}

The site is still in draft mode of what we want it to be, but playing with it, we can already see some uses for those looking to learn about subjects related to the listed entities. 
In Figure 12 we see the website running, after receiving the data from the dygiepp-reader API. 
In this case, the entity "decentralized application" was selected on the left, and then all the relations of which the word or any of its aliases are part were listed. The use of aliases can be seen due to the fact that we have relationships for "dApps" and "decentralized app", both aliases of "decentralized application". Some improvements can be made, like specifying to the API which dataset we would like to get information from in a call. At the moment, the API only serves data extracted from the whitepaper dataset, but could be updated to serve both the corpus of academic articles and Wikipedia texts. It would also be interesting to combine the results, joining the data obtained from the three datasets. When you click on a relationship, the phrase it was taken from is displayed, but we would also like to know which document and in which phrase number it was taken. There are also some easily solvable problems such as the presence of repeated relationships. 

\section{Conclusions}
The development of this project was inspired by the work of Yi Luan et al~\cite{luan2018, luan-etal-2019-general}, where the SciIE framework for extracting entities and relationships from a dataset was developed. This was meant to produce entities and relationships, organized into a knowledge graph, to be used for organizing information from a  domain-specific dataset, making it an important tool for research purposes. 
Then came the hypothesis that this model of knowledge graph construction could be applied not only to the area of artificial intelligence as it was in the case of SciIE, which uses the dataset related to AI, the SciERC, but also to the blockchain area.
If the article introducing SciIE helped to formulate the initial idea for this project, the work of David Wadden et al~\cite{wadden-etal-2019}, which introduces the DyGIE++ framework, was essential for its realization. DyGIE++ was developed by some of the same authors as SciIE. This program lowered the entry barrier for the application of prediction models to perform the extraction of entities and relationships, a fundamental task in the construction of the knowledge graphs that this project sought to build. 

It is worth mentioning again the use of the AllenNLP framework, which provided commands via the command line that were fast and intuitive, allowing an undergraduate project to make use of concepts that would generally be limited to more advanced work, such as a master's or doctoral studies.
As seen in the analysis of the results obtained  and with the visualization of the complete knowledge graph, the hypothesis that SciERC could be used with DyGIE++ to extract entities and relationships in blockchain-related data seems to be correct. With this in mind, we suggest that this work can be applied to other areas in the computing domain, whether they are more established or newer areas such as the blockchain. The application of this technique of knowledge graph construction to other areas of computing should bring important developments both to the areas in question and for the technique itself.

Much of the motivation for this project is due to the fact that the blockchain area is so new, with few useful tools to support research on emerging projects and technologies in this domain. Most of the available tools are aimed at analyzing data related to blockchain statistics (number of transactions, value of fees, etc.), cryptocurrency price movements and technical analysis of price charts~\cite{kibet}. The lack of  tools that allow the visualization of the blockchain ecosystem  (not limited to statistics and market data) was seen as an opportunity to develop a project that was relevant in this new and innovative domain.

Although the development of this work did not require the construction of new code to perform the tasks of extracting entities and relations, knowledge of data structures, Javascript and web development was necessary to deliver the products of this work. Dygiepp-reader was developed, a Javascript program for manipulating and structuring DyGIE++ results, as well as a web interface to visualize the obtained results. As an API, the dygiepp-reader can even evolve to have routes that accept parameters, such as the data corpus for which the client wants to obtain the results. It is also interesting that this API should be able to combine the results of different corpora of information in order to create a unified knowledge graph. The web interface can also evolve to become more interactive, letting the user choose exactly which document and which
phrases an entity or relationship has. 

As this work has a special focus on collecting data related to blockchain projects, it would be interesting to allow the user to search for the entities and relationships of a specific project.
Both products, the dygiepp-reader program and the web interface will be made available to the public. 

Finally, we note that this report is the abridged English version of the undergraduate thesis \textit{Visualizando o Ecossistema Blockchain Utilizando
Machine Learning e Grafos de Conhecimento} of the first author, jointly supervised by the two other authors.


\addtocontents{toc}{\protect\setcounter{tocdepth}{1}}
\addcontentsline{toc}{section}{References}
\bibliography{references.bib}{}
\bibliographystyle{acm}

\pagebreak

\end{document}